\documentclass[english]{article}
\usepackage{graphicx,color}
\def\no{\noindent}
\def\bc{\begin{center}}
\def\ec{\end{center}}

\def\beq{\begin{equation}}
\def\eeq{\end{equation}}

\def\br{{\bf r}}
\def\bq{{\bf q}}

\begin{document}

\title{Quantum diffusion in two-dimensional random systems with particle-hole symmetry
}

\author{K. Ziegler\\
Institut f\"ur Physik, Universit\"at Augsburg\\
D-86135 Augsburg, Germany}

\maketitle

\no
Abstract:

\no
We study the scattering dynamics of an $n$-component spinor wavefunction in a random environment
on a two-dimensional lattice. If the particle-hole symmetry of the Hamiltonian is spontaneously broken
the dynamics of the quantum particles becomes diffusive on large scales. The latter is described by 
a non-interacting Grassmann field, indicating a special kind of asymptotic freedom on large scales in $d=2$.
\vskip0.5cm

\no


\section{Introduction}

Conventional wisdom is that a classical approach of a randomly scattered
particle leads to diffusion. Diffusion in quantum systems can either be 
caused by particle-particle collisions or collisions with (static) impurity scatterers. 
If the latter are randomly distributed, however, this may lead to Anderson localization rather
than to diffusion \cite{anderson58,abrahams79}. This effect is particularly strong in 
low-dimensional systems such as in two-dimensional graphene sheets.
The scaling approach to generic random scattering \cite{abrahams79}
indicates that diffusion is suppressed by Anderson localization for dimension $d\le 2$.
On the other hand, ballistic motion can also be ruled out, even for a finite system with random scattering
\cite{simon90}. It was pointed out by Kaveh, however, that diffusion cannot be obtained in random-phase
approximation applied to t he disordered system \cite{kaveh85}.

Inspired by the recent observation of metallic behavior (i.e. diffusive or even ballistic transport) in
disordered two-dimensional systems (graphene) 
\cite{novoselov05,zhang05}, a general discussion of a diffusive quantum particle is required, which takes into
account a spinor structure of the wavefunction.
There are two possibilities, ballistic transport for finite systems \cite{rosenstein10,morandi11} 
or diffusive transport for infinite systems \cite{ziegler97}.
Here we will focus on infinite systems and study a quantum $n$--component
spinor particle on a two-dimensional lattice with particle-hole symmetry. It will be shown that 
on large scales the particle diffuses on the lattice with $d=2$. 
This work presents a generalization of the idea that a spontaneously broken
supersymmetry can lead to diffusion in a system with particle-hole symmetry
\cite{ziegler97,Ziegler2009}.

The motion of a quantum particle is characterized by the transition probability $P_{\br,\br'}(i\epsilon)$ for an 
$n$--component spinor particle at site $\br'$ that moves to site $\br$ with frequency $i\epsilon$:
\beq
P_{\br,\br'}(i\epsilon)=\frac{K_{\br,\br'}(i\epsilon)}{\sum_{\br}K_{\br,\br'}(i\epsilon)}\ \ \ {\rm with}\ \ \
K_{\br,\br'}(i\epsilon)=\langle {\rm Tr}_n\left[ G_{\br,\br'}(i\epsilon)
G^\dagger_{\br',\br}(i\epsilon)\right]\rangle_v
=\langle {\rm Tr}_n\left[ G_{\br,\br'}(i\epsilon)G_{\br',\br}(-i\epsilon)\right]\rangle_v 
\ ,
\label{transition00}
\eeq
where $G(i\epsilon)=(i\epsilon+H)^{-1}$ is the one-particle Green's function of the Hamiltonian $H$
and $\langle ...\rangle_v$ is the average with respect to some random scatterers. ${\rm Tr}_n(...)$ is the trace
with respect to the $n$ spinor components. The last equation in Eq. 
(\ref{transition00}) follows from the Hermitean Hamiltonian: $H^\dagger=H$.

After Fourier transformation of the two-particle Green's function $K_{\br,\br'}(i\epsilon)\to k_{\br,\br'}(t)$ we 
study the motion of the quantum particle with the mean-square displacement
of the coordinate $r_k$
\beq
\langle r_k^2\rangle 
= \frac{\sum_\br r_k^2 k_{\br,0}(t)}{\sum_\br k_{\br,0}(t)}
\ .
\label{diff2}
\eeq
This expression grows linearly with time $t$ in the case of
diffusion.


\section{Model}

We consider an $n$-component spinor wavefunction 
described by the Hamiltonian matrix
\beq
H=H_0+vH_1
\ , \ \ \ H_0=(h_{\br,\alpha;\br',\alpha'})
\ , \ \ \ H_1=(h_{\alpha,\alpha'}\delta_{\br,\br'})
\ , \ \ \ v=(v_{\br}\delta_{\alpha,\alpha'}\delta_{\br,\br'})
\ ,
\label{hamilt00}
\eeq
where $\br,\br'$ are coordinates on the two-dimensional lattice and $\alpha,\alpha'=1,2,...,n$
refer to the $n$ spinor components. $v_\br$ is a random variable with an uncorrelated Gaussian distribution:
$\langle v_\br\rangle_v=0$, $\langle v_\br v_{\br'}\rangle_v=g\delta_{\br,\br'}$.
In the following we assume that the Hamiltonian satisfies the generalized particle-hole symmetry
$
H_j\to -UH_j^*U^\dagger =H_j
$  (j=0,1), which belongs to class D according to Cartan's classification scheme \cite{zirnbauer96}. 
In terms of the Green's functions, this transformation provides a sign change of the
frequency: 
\beq
G(i\epsilon)\to -UG^T(i\epsilon)U^\dagger=G(-i\epsilon)
\ ,
\label{gf_relation}
\eeq
since $H_j^\dagger =H_j$ implies $H_j=-UH_j^TU^\dagger$ ($^T$ is the matrix transposition).
The Green's functions $G(i\epsilon)$ and the transposed Green's function
$G^T(i\epsilon)$ can be expressed in a functional-integral representation of a free field complex (boson) field 
$\phi^1_{\br,k}$ and a Grassmann (fermion) field $\phi^2_{\br,k}$, respectively. This allows us to construct
the Bose-Fermi functional integral \cite{negele}
\beq
\langle f(\phi)\rangle_\phi
=\int f(\phi) e^{-S}{\cal D}[\phi]
\label{functint0}
\eeq
which is normalized:
\beq
\int e^{-S}{\cal D}[\phi] = 1
\ .
\label{norm}
\eeq
The action $S$ is
\beq
S=-i(\phi\cdot({\hat H}_0+i\epsilon){\bar \phi})+g(\phi\cdot {\hat H}_1{\bar \phi})^2 \ \ \ 
(\epsilon > 0)
\ ,
\label{action0}
\eeq
with respect to the boson-fermion vector field $\phi=(\phi^1_{\br,k},\phi^2_{\br,k})$ ($k=1,2,...,n$) and with 
the block-diagonal Hermitean matrices ${\hat H}_j=diag(H_j,H_j^T)$.
After averaging over the random variables $v_\br$ we can write 
\beq
\langle G_{\br,k;\br',l}(i\epsilon)G^T_{\br',m;\br,n}(i\epsilon)\rangle_v
=-\langle\phi_{\br',l}^1{\bar\phi}_{\br,k}^1\phi_{\br,n}^2{\bar\phi}_{\br',m}^2\rangle_\phi
\label{expect1}
\eeq
with $\langle ...\rangle_\phi = \int ... e^{-S}{\cal D}[\phi]$. 
The normalization can easily be seen by performing the $\phi$ integration before averaging over $v_\br$.

An integral of the form (\ref{norm}) describes a supersymmetric field theory, meaning that it is a field theory
for bosons as well as fermions which appear with the {\it same} Green's functions \cite{negele}. 
However, it should be noticed that supersymmetry is sufficient for the normalized integral but not necessary
\cite{ziegler97}. 
In the present case the boson and the fermion Green's functions are different, provided that $H^T\ne H$.
The choice of different Green's functions in the action (\ref{action0}) has profound consequences in comparison 
with the model, where fermions and bosons appear symmetrically with the same Green's function, because the
latter is subject to a larger symmetry group. This will be discussed at the end of the paper.

Using the relation in Eq. (\ref{expect1}), we can write for the expression in Eq. (\ref{transition00}) as
\[
K_{\br,\br'}=\langle {\rm Tr}_n\left[ G_{\br,\br'}(i\epsilon)G_{\br',\br}(-i\epsilon)\right]\rangle_v
=-\langle {\rm Tr}_n\left[ G_{\br,\br'}(i\epsilon)UG^T_{\br',\br}(i\epsilon)U^\dagger\right]\rangle_v
\]
\beq
=\sum_{l,m,n,n'}U_{m,n}U_{l,n'}^*
\langle\phi_{\br',m}^1{\bar\phi}_{\br,l}^1\phi_{\br,n'}^2{\bar\phi}_{\br',n}^2\rangle_\phi
=-\sum_{l,m,n,n'}U_{m,n}U_{l,n'}^*
\langle\phi_{\br',m}^1{\bar\phi}_{\br',n}^2\phi_{\br,n'}^2{\bar\phi}_{\br,l}^1\rangle_\phi
\ .
\label{trans3}
\eeq
This expression will be used subsequently to study diffusion in the particle-hole symmetric system.

\section{Summary of the subsequent calculation}

Before embarking to the detailed calculation of $K_{\br,\br'}$, that will lead us to a simple expression for 
the functional integral on large scales $|\br-\br'|$ in terms of a saddle-point approximation, 
a brief outlook on the lengthy calculation is given in this section. In a first step we identify a symmetry
in terms of a similarity transformation with respect to the boson-fermion structure. After introducing 
a new field in the functional integral, we apply a saddle-point approximation to the latter. It turns out that
the above mentioned symmetry creates a two-dimensional fermionic saddle-point manifold, given by a two-component
Grassmann field $(\varphi,\varphi')$. For large scales this becomes a free field and provides a diffusion propagator.
In other words, our approximation scheme allows us to prove that the Fourier components of $K_{\br,\br'}(i\epsilon)$
describe diffusion in the large distance asymptotics:
\beq
K_\bq(i\epsilon)\sim\frac{{\bar K}}{ib\epsilon+{\tilde c}_0-{\tilde c}_\bq}
\label{diff_prop0}
\eeq  
with finite constants $b$ and
$
{\bar K}
$, which is determined by the solution of the saddle-point equation.
Moreover, ${\tilde c}_\bq$ are the Fourier components of
\[
c_{\br,\br'} 
=16 {\rm Tr}_n\left[g_{+,\br',\br}Q_2H_1g_{-,\br,\br'}Q_2H_1\right]
\ ,
\]
where the Green's functions $g_\pm$ are defined as
\beq
g_\pm=[H_0\pm i\epsilon+2(Q_1\pm Q_2)H_1]^{-1}
\ .
\label{green4}
\eeq
$Q_1$, $Q_2$ are determined by saddle-point equations.

\no
{\it remark}: The Green's functions $g_\pm$ can be considered as the self-consistent Born approximation (SCBA) 
of the random Green's functions $G(\pm i\epsilon)$, where $Q_1$, $Q_2$ are self-energies \cite{abergel10,ando98}. 

\section{Diffusion on large scales}

In the following we derive the asymptotic form of $K_{\br,\br'}$ in Eq. (\ref{diff_prop0}).

\subsection{Boson-Fermion Symmetry}

Considering the block matrix
\[
\pmatrix{
A & \Theta \cr
{\bar \Theta} & B \cr
}
\ ,
\]
where the elements of the matrices $A$, $B$ are complex and the elements of the
matrices $\Theta$, ${\bar\Theta}$ are Grassmannian, we introduce the graded trace 
\[
{\rm {\rm Tr}g}\pmatrix{
A & \Theta \cr
{\bar \Theta} & B \cr
}={\rm Tr} A - {\rm Tr} B \ ,
\]
where ${\rm Tr}$ is the conventional trace, and the graded determinant ${\rm detg}$ \cite{ziegler97}:
\beq
{\rm detg} \pmatrix{
A & \Theta \cr
{\bar \Theta} & B \cr
}=\frac{\det(A)}{\det(B)}\det({\bf 1}-\Theta B^{-1}{\bar\Theta} A^{-1}) 
=\frac{\det(A-\Theta B^{-1}{\bar\Theta})}{\det(B)}
\ .
\label{detg}
\eeq
For the special matrix
$
{\hat H}=diag(H,H^T)
$
this gives ${\rm Trg}({\hat H})=0$ and ${\rm detg}({\hat H}+i\epsilon)=1$.
${\rm Trg}$ and ${\rm detg}$ have the same properties as the conventional trace and determinant.
In particular, we have the relations ${\rm detg}({\hat A}){\rm detg}({\hat B})={\rm detg}({\hat A}{\hat B})$
and ${\rm detg}({\hat A})=\exp({\rm Trg}(\log {\hat A}))$.

Now we consider the special matrix
\beq
{\hat S}=\pmatrix{
0 & \varphi U \cr
\varphi' U^\dagger & 0\cr
} \ \ \ (\varphi, \varphi' \in G)
\ ,
\label{S}
\eeq
where $G$ is a Grassmann algebra (i.e. $\varphi\varphi'=-\varphi'\varphi$).
${\hat H}_j$ and ${\hat S}$ anticommute:
\[
{\hat H}_j{\hat S}=\pmatrix{
0 & H_j\varphi U \cr
H_j^T\varphi' U^\dagger & 0 \cr
}=\pmatrix{
0 & \varphi H_jU \cr
\varphi' H_j^TU^\dagger & 0 \cr
}
=\pmatrix{
0 & -\varphi UH_j^T \cr
-\varphi' U^\dagger H_j & 0 \cr
}=-{\hat S}{\hat H}_j
\ ,
\]
where the second equation follows from the assumption that $\varphi$, $\varphi'$
commute with $H_j$. 
This relation implies that for a global ${\hat S}$ (i.e., ${\hat S}$ is constant on the lattice)
\beq
e^{\hat S} {\hat H}_je^{\hat S}={\hat H}_j
\ ,
\label{symm3}
\eeq
which can be considered the supersymmetry of the model defined in (\ref{action0}) because the 
transformation connects the fermionic and the bosonic sector of the theory. 
For the subsequent calculations it is useful to notice that with ${\rm Trg} {\hat S}=0$ we have 
\beq
{\rm detg}(e^{\hat S})=\exp \left( {\rm Trg} {\hat S}\right) = 1
\ .
\label{invariance1}
\eeq

\subsection{Functional integral with nonlinear field}

Defining the tensor field
\[
{\hat \Phi}^{j,j'}_{kk'}={\bar\phi}^{j}_k \phi^{j'}_{k'}  \ \ \
(j,j'=1,2;\ \ k,k'=1,...,n)
\ ,
\]
we rewrite the terms in Eq. (\ref{action0}) as
\[
(\phi\cdot {\hat H}_0{\bar \phi})={\rm Trg}({\hat H}_0{\hat \Phi}) , \ \ \ 
(\phi\cdot {\hat H}_1{\bar \phi})^2={\rm Trg} ({\hat H}_1{\hat \Phi}{\hat H}_1{\hat \Phi})
\ .
\]
Then the identity
\[
g{\rm Trg} ({\hat H}_1{\hat \Phi}{\hat H}_1{\hat \Phi})
+g^{-1}{\rm Trg}[(ig{\hat H}_1{\hat \Phi}-{\hat Q})(ig{\hat H}_1{\hat \Phi}-{\hat Q})]
=g^{-1}{\rm Trg}({\hat Q}^2)-2i{\rm Trg}({\hat Q}{\hat H}_1{\hat \Phi})
\]
with matrix field
\[
{\hat Q}=\pmatrix{
Q_\br & \Theta_\br \cr
{\bar\Theta}_\br & iP_\br \cr
}
\]
allows us to write the interaction term as a ${\hat Q}$ integral:
\[
\exp\left[-g{\rm Trg} ({\hat H}_1{\hat \Phi}{\hat H}_1{\hat \Phi})\right]
=\int\exp\left[-g{\rm Trg} ({\hat H}_1{\hat \Phi}{\hat H}_1{\hat \Phi})
-g^{-1}{\rm Trg}[(ig{\hat H}_1{\hat \Phi}-{\hat Q})(gi{\hat H}_1{\hat \Phi}-{\hat Q})]\right]
{\cal D}[{\hat Q}]
\]
\[
=\int\exp\left[-g^{-1}{\rm Trg}({\hat Q}^2)+2i{\rm Trg}({\hat Q}{\hat H}_1{\hat \Phi})\right]
{\cal D}[{\hat Q}]
\ .
\]
With the expression on the right-hand side we can
perform the $\phi$ integration in the functional integral of Eq. (\ref{functint0}), 
since $\phi$ appears only as a quadratic form in the exponent. Thus we remain with a functional integral 
over ${\hat Q}$:
\beq
\int F({\hat \Phi}) e^{-S}{\cal D}[\phi]
=\int G({\hat Q}){\rm detg}({\hat H}_0+i\epsilon +2{\hat Q}{\hat H}_1)^{-1} 
e^{-g^{-1} {\rm Trg}({\hat Q}^2)}{\cal D}[{\hat Q}]
\label{functint2}
\eeq
because of
\[
\int e^{-{\rm Trg}({\hat A}{\hat\Phi})}{\cal D}[\phi] = {\rm detg}({\hat A})^{-1}
\ .
\]
The determinant
\beq
J={\rm detg}({\hat H}_0+i\epsilon +2{\hat Q}{\hat H}_1)^{-1} 
\label{jacobian2a}
\eeq
is the Jacobian for the transformation $\phi\to {\hat Q}$ in the functional integration.
The function $G$ can be obtained from $F$ by directly calculating the integrals on both
sides. This, however, is a complex task for a general $F$. Here we consider only one
specific case which is sufficient for a diffusive mode:
\[
K_{\br,\br'}=-\frac{1}{g^2} 
\sum_{l,m,n,n'}U_{m,n}U_{l,n'}^*
\langle( H_1^{-1}\Theta_{\br'})_{mn}( H_1^{T-1}{\bar\Theta}_\br)_{n'l}\rangle_{\hat Q}
\]
with $\langle ... \rangle_{\hat Q}=\int ... Je^{-g^{-1} {\rm Trg}({\hat Q}^2))}{\cal D}[{\hat Q}]$.
Moreover, we have $H_1^T=-U^\dagger H_1U$ such that
\beq
K_{\br,\br'}=\frac{1}{g^2} 
\sum_{l,m,n,n'}U_{m,n}U_{l,n'}^*
\langle( H_1^{-1}\Theta_{\br'})_{mn}( U^\dagger H_1^{-1}U{\bar\Theta}_\br)_{n'l}\rangle_{\hat Q}
\ .
\label{diff2a}
\eeq

\subsection{Saddle-point approximation}

The saddle-point approximation of the functional integral (\ref{functint2}) is given
by a solution of the saddle-point equation $\delta_{\hat Q} S'=0$ with
\beq
S'= g^{-1} {\rm Trg}({\hat Q}_0^2)+\log{\rm detg}\left( {\hat H}_0+i\epsilon +2{\hat Q}{\hat H}_1 \right)
\ .
\label{action3}
\eeq
The saddle-point is degenerate with respect to the similarity transformation
\[
e^{{\hat S}}{\hat Q}_0e^{-{\hat S}} \ \ \ {\rm with} \ \ 
{\hat Q}_0=\pmatrix{
Q_0 & 0\cr
0 & iP_0 \cr
}
\ .
\]
This covers the entire saddle-point degeneracy because we assume here that there is no additional symmetry
of $H$.

${\hat Q}_0$ consists of two terms, namely ${\hat Q}_0={\hat Q}_1+{\hat Q}_2$,
where ${\hat Q}_1$ (${\hat Q}_2$) commutes (anticommutes) with ${\hat S}$:
\beq
e^{{\hat S}}{\hat Q}_0 e^{-{\hat S}}={\hat Q}_1+{\hat Q}_2e^{-2{\hat S}}
\label{transf3}
\eeq
which implies ${\rm Trg}({\hat Q}_0^2)=0$. Then the saddle-point solution contributes to the action (\ref{action3})
the two terms
\beq
U(Q_1H_1)^TU^\dagger=-Q_1H_1, \ \ \  U(Q_2H_1)^TU^\dagger=Q_2H_1
\ ,
\eeq
where the first (second) term preserves (breaks) the symmetry of the Jacobian.
These properties imply 
\beq
Ug_-^TU^\dagger=-g_+
\label{gf_relation1}
\eeq
 which is consistent with Eq. (\ref{gf_relation}).

Inserting the expression (\ref{transf3}) into the functional integral of Eq. (\ref{functint2}) results in 
\beq
\int G(e^{{\hat S}}{\hat Q}_0e^{-{\hat S}}) 
{\rm detg}
\left[ {\hat H}_0+i\epsilon +2{\hat Q}_1{\hat H}_1+2{\hat Q}_2{\hat H}_1e^{2{\hat S}} \right]^{-1}
{\cal D}[{\hat Q}']
\ .
\label{fint5}
\eeq
This indicates that ${\hat Q}_2$ is the order parameter for spontaneous symmetry breaking.
Thus we have reduced the integration to the nonlinear field ${\hat Q}'={\hat Q}_2\exp(2{\hat S})$, 
while ${\hat Q}_0$ is determined by the saddle-point condition.

Now we use the identity $e^{2{\hat S}}=2({\bf 1}-{\hat S})^{-1}-{\bf 1}$ and define 
$
\gamma_\pm=4g_\pm Q_2H_1
$
with the help of the Green's functions $g_\pm$ in Eq. (\ref{green4}) 
to obtain for the inverse Jacobian (cf. Appendix \ref{app:jacobian})
\beq
J^{-1}={\bar J}^{-1} 
\det({\bf 1}+\gamma_+\varphi \varphi'-\varphi \gamma_-\varphi'+\gamma_+\varphi \gamma_-\varphi') \ \ \ 
{\rm with}\ \ 
{\bar J}=\frac{\det(-\left[H_0-i\epsilon+2(Q_1-Q_2)H_1\right])}{\det(H_0+i\epsilon+2Q_0H_1)}
\ .
\label{jacobian2a}
\eeq
Using the identity 
$\det(A)=\exp\{{\rm Tr}[\log(A)]\}$, we eventually have
\beq
J={\bar J}\exp\left\{-{\rm Tr}\left[\log\left({\bf 1}+\gamma_+\varphi\varphi' 
- \varphi \gamma_-\varphi' +\gamma_+\varphi \gamma_-\varphi'\right)\right]
\right\}
\ .
\label{jacobian2b}
\eeq

\subsection{Large-scale properties}
\label{sect:diff}

The spatial diagonal elements of
$\gamma_+\varphi\varphi' - \varphi \gamma_-\varphi' +\gamma_+\varphi \gamma_-\varphi'$
can be written as
\beq
(\gamma_+\varphi\varphi' - \varphi \gamma_-\varphi' +\gamma_+\varphi \gamma_-\varphi')_{\br,\br}
=(\gamma_+-\gamma_- +\gamma_+ \gamma_-)_{\br,\br}\varphi_\br\varphi'_\br
+\sum_{\br'}\gamma_{+,\br,\br'} \gamma_{-,\br',\br}(\varphi_{\br'}-\varphi_\br)\varphi'_\br
\ ,
\label{matrix11}
\eeq
where the first part is proportional to $\epsilon$:
\beq
(\gamma_+-\gamma_- +\gamma_+ \gamma_-)_{\br,\br}
=-8i\epsilon g_+g_-Q_2H_1
\ .
\label{symm_break}
\eeq
The second term can also be expressed as
\beq
\sum_{\br'}\gamma_{+,\br,\br'} \gamma_{-,\br',\br}(\varphi_{\br'}-\varphi_\br)\varphi'_\br
=\sum_{\br'}d_{\br,\br'}\varphi_\br\varphi_{\br'}'
\label{corr2}
\eeq
with
\beq
d_{\br,\br'}=\delta_{\br,\br'}\sum_{\br''}c_{\br'',\br'}-c_{\br,\br'} \ \ \ {\rm with}\ \ \ 
c_{\br,\br'}= {\rm Tr}_n\left[\gamma_{+,\br',\br}\gamma_{-,\br,\br'}\right]
\ .
\label{corr8}
\eeq
It should be noticed in Eq. (\ref{corr2}) that the spatial diagonal elements 
$\gamma_{\pm,\br,\br}$ do not contribute. Moreover,
$(\gamma_+\varphi\varphi' - \varphi \gamma_-\varphi' +\gamma_+\varphi \gamma_-\varphi')_{\br,\br'}$
($\br'\ne \br$) has at least one spatial off-diagonal factor $\gamma_{\pm,\br,\br'}$ in each term.
Therefore, all matrix elements
$(\gamma_+\varphi\varphi' - \varphi \gamma_-\varphi' +\gamma_+\varphi \gamma_-\varphi')_{\br,\br'}$
have at least one  factor $\gamma_{\pm,\br,\br'}$ with $\br'\ne \br$, except for 
the diagonal term in (\ref{symm_break}) which is proportional to $\epsilon$.

In the next step we analyze terms that depend on the off-diagonal elements $\gamma_{\pm,\br,\br'}$
($\br'\ne \br$).
Under a change of the length scale $\br\to \Delta \br$ on the two-dimensional lattice
these off-diagonal terms scale as (cf. Appendix \ref{app:scaling})
\beq
\gamma_{\pm,\br,\br'}\to \Delta^{-2}\gamma_{\pm,\br,\br'} \ \ \ (\br'\ne \br)
\ .
\label{scaling1}
\eeq
$\epsilon$ is an arbitrarily small parameter which should be sent to zero. This allows
us to replace $\epsilon\to\Delta^{-2}\epsilon$ here.
Moreover, products of $n$ matrices are of order $\Delta^{-2n}$ because $\gamma_{\pm,\br,\br'}$
decays exponentially in space due to the nonzero symmetry breaking term $Q_2$. Therefore, 
the intermediate $\br$ summations do not contribute a factor $\Delta$. 
Finally, the trace scales as ${\rm Tr}\to\Delta^2 {\rm Tr}$, and we obtain from Eq. (\ref{jacobian2b}) 
for the scaled Jacobian
\[
J\to J_\Delta
={\bar J}_\Delta\exp\left\{-\Delta^2
{\rm Tr}\left[\log\left({\bf 1}+\Delta^{-2}(\gamma_+\varphi\varphi' 
- \varphi \gamma_-\varphi' +\gamma_+\varphi \gamma_-\varphi')\right)
\right]\right\}
\ .
\]
Thus the large-scale limit $\Delta\sim\infty$ reads
\beq
J_\Delta\sim
{\bar J}_\Delta\exp\left\{-{\rm Tr}\left(
\gamma_+\varphi\varphi' - \varphi \gamma_-\varphi' +\gamma_+\varphi \gamma_-\varphi'
\right)\right\}
\ ,
\label{asymptotics}
\eeq
which is a quadratic form of $\varphi$, $\varphi'$ in the exponent (i.e., $(\varphi,\varphi')$ is a free field). 
This reads with Eqs. (\ref{symm_break}), (\ref{corr2})
\beq
J_\Delta\sim {\bar J}_\Delta\exp \left[-\sum_{\br,\br'}
\left(i\epsilon b\delta_{\br,\br'}+d_{\br,\br'}\right)\varphi_\br\varphi_{\br'}'\right]
\equiv{\bar J}_\Delta\exp\left(-\sum_{\br,\br'}\kappa^{-1}_{\br,\br'}\varphi_\br\varphi_{\br'}'\right)
\ ,
\label{jacobian5}
\eeq
where
$
b=8{\rm Tr}_n[(g_+g_-Q_2H_1)_{\br,\br}]
$.
After Fourier transformation $\br\to\bq$ we obtain
\beq
{\tilde d}_\bq={\tilde c}_0-{\tilde c}_\bq  \ \ \ {\rm and} \ \
\kappa_\bq=\frac{1}{ib\epsilon+{\tilde c}_0-{\tilde c}_\bq}
\ .
\label{FT_kappa}
\eeq
Returning to the functional integral in Eq. (\ref{diff2a}) we now have an integration 
over $\varphi$, $\varphi'$ with
\[
\Theta_\br=-2Q_2U\varphi_\br , \ \ \
{\bar \Theta}_\br=2U^\dagger Q_2\varphi_\br'
\]
such that
\[
K_{\br,\br'}\sim \frac{4{\bar J}_\Delta}{g^2} 
\sum_{m,n}U_{m,n}( H_1^{-1}Q_2U)_{mn}\sum_{l,n'}U_{l,n'}^*( U^\dagger H_1^{-1}Q_2)_{n'l}
\langle \varphi_\br\varphi_{\br'}'\rangle
\]
\[
=\frac{{4\bar J}_\Delta}{g^2}{\rm Tr}_n(UU^TH_1^{-1}Q_2){\rm Tr}_n(U^*U^\dagger H_1^{-1}Q_2)
\langle \varphi_\br\varphi_{\br'}'\rangle
\]
with $\langle \varphi_\br\varphi_{\br'}'\rangle =-\kappa_{\br',\br}/\det(\kappa)$. 
Using the Fourier components in Eq. (\ref{FT_kappa}), the Fourier transformation of $K_{\br,\br'}$ reads
\beq
{\tilde K}_\bq\sim\frac{{\bar K}}{ib\epsilon+{\tilde c}_0-{\tilde c}_\bq}
\ ,
\label{diffusion_prop}
\eeq
where
\[
{\bar K}
=\frac{4{\bar J}_\Delta}{\det(\kappa)g^2}{\rm Tr}_n(UU^TH_1^{-1}Q_2){\rm Tr}_n(U^*U^\dagger H_1^{-1}Q_2)
\ .
\] 
This concludes our calculation of the large-scale properties of $K_{\br,\br'}$.

\subsection{Alternative approach: Nonlinear sigma model}

Returning to the expression in Eq. (\ref{fint5}), we can expand the logarithm of the Jacobian in powers 
of ${\hat Q}_2$ up to second order. This approximation is referred to as the nonlinear sigma model approach 
which is believed to provide a good description of the transport properties of disordered systems 
\cite{wegner80,efetov97}.
For our model we derive the nonlinear sigma model for the action
\[
S'=\log\left[{\rm detg}
\left( {\hat H}_0+i\epsilon +2{\hat Q}_1{\hat H}_1+2{\hat Q}_2{\hat H}_1e^{2{\hat S}} \right)
\right]
\]
\[
=\log\left[{\rm detg}
\left( {\hat H}_0+i\epsilon +2({\hat Q}_1+{\hat Q}_2){\hat H}_1
+2{\hat Q}_2{\hat H}_1(e^{2{\hat S}}-{\bf 1}) \right)\right]
\ ,
\]
where $e^{2{\hat S}}-{\bf 1}=2({\hat S}+{\hat S}^2)$. With
\[
{\hat G}_0=\pmatrix{
g_+ & 0 \cr
0 & -U^\dagger g_-U \cr
}^{-1}=\pmatrix{
g_+ & 0 \cr
0 & g_+^T \cr
}^{-1}
\]
we can expand the action up to second order in ${\hat Q}_2$ as $S'\approx S_0+S''$ with
\[
S''=4{\rm Trg}\left({\hat G}_0{\hat Q}_2{\hat H}_1({\hat S}+{\hat S}^2)\right)
+8 {\rm Trg}\left[\left({\hat G}_0{\hat Q}_2{\hat H}_1({\hat S}+{\hat S}^2)\right)^2\right]
\]
\beq
=4{\rm Trg}\left({\hat G}_0{\hat Q}_2{\hat H}_1{\hat S}^2\right)
+8 {\rm Trg}\left[\left({\hat G}_0{\hat Q}_2{\hat H}_1{\hat S}\right)^2\right]
+8 {\rm Trg}\left[\left({\hat G}_0{\hat Q}_2{\hat H}_1{\hat S}^2\right)^2\right]
\ .
\label{expansion2}
\eeq
${\hat G}_0{\hat Q}_2{\hat H}_1$ can be approximated by a gradient operator. This gives
the standard form of the nonlinear sigma model for the last two terms, whereas the first term
contributes to the symmetry-breaking term which is proportional to $i\epsilon$. 
Moreover, a straightforward calculation shows that the last term vanishes for our model
\beq
{\rm Trg}\left[\left({\hat G}_0{\hat Q}_2{\hat H}_1{\hat S}^2\right)^2\right]=0
\ ,
\label{nonlinear1}
\eeq
such that only the quadratic terms in $\varphi$ survive in the nonlinear sigma model. This is in agreement
with the exponent in Eqs. (\ref{asymptotics}) and (\ref{jacobian5}).

\section{Discussion}

Our derivation of $K_{\br,\br'}$ in the previous section was obtained without 
specifying $H_0$, $H_1$ of the Hamiltonian. This prevents us from determining $Q_{1}$, $Q_{2}$
here because this requires the solution of the saddle-point equation. It is crucial though that
the symmetry breaking term $Q_2$ represents a mass to the Green's functions $\gamma_\pm$ such that
the latter decay exponentially. There is no diffusion but localization
for saddle-point solutions with $Q_2=0$, as discussed for the case of Weyl fermions in Ref. \cite{Ziegler2009}. 

We leave the determination of $Q_{1}$, $Q_{2}$ for specific 
Hamiltonians to further work and study only the 
general structure of the diffusion propagator in Eq. (\ref{diffusion_prop}). 
For the large-scale behavior of the latter we consider $q\sim 0$
\beq
{\tilde K}_\bq
\sim\frac{{\bar K}}{b}\frac{1}{i\epsilon+ \sum_{i,j}D_{ij} q_i q_j}
\label{diffusion_prop2}
\eeq
with 
\[
{\tilde d}_\bq={\tilde c}_0-{\tilde c}_\bq\sim b\sum_{i,j} D_{ij}q_i q_j 
\]
and with the diffusion coefficients
\[
D_{ij}=-\frac{1}{2b}\frac{\partial^2{\tilde c}_q}{\partial q_i\partial q_j}\Big|_{q=0} 
=\frac{\sum_\br r_ir_j{\rm Tr}_n\left[g_{+,0,\br}Q_2H_1g_{-,\br,0}Q_2H_1\right]}
{{\rm Tr}_n[(g_+g_-Q_2H_1)_{\br,\br}]}
\ .
\]
In the isotropic case (i.e. for $D_{ij}=D\delta_{ij}$) we have
\[
{\tilde d}_\bq={\tilde c}_0-{\tilde c}_\bq\sim bDq^2 , \ \ \
D=-\frac{1}{2b}\frac{\partial^2{\tilde c}_\bq}{\partial q_k^2}\Big|_{q=0} 
=\frac{1}{2b}\sum_\br r_k^2 c_{\br,0}
\]
such that the diffusion propagator reads
\beq
{\tilde K}_\bq(i\epsilon)=\frac{{\bar K}}{ib\epsilon+{\tilde c}_0-{\tilde c}_\bq}
\sim\frac{{\bar K}}{b}\frac{1}{i\epsilon+ D q^2}
\ .
\label{diffusion_prop1}
\eeq
From the diffusion propagator we can evaluate the dynamics of the quantum walk. We apply 
a Fourier transformation from frequency $\epsilon$ to time $t$ and get
\[
{\tilde K}_\bq(i\epsilon)\to K_\bq(t)= \frac{{\bar K}}{b} e^{-Dq^2t}
\ ,
\]
and a Fourier transformation from momentum $\bq$ to real space coordinates $\br$ gives
\[
K_\bq(t)\rightarrow k_\br(t)=\frac{{\bar K}}{b}\frac{e^{-r^2/4Dt}}{\pi Dt}
\ .
\]
This provides the mean-square displacement as a function of time:
\beq
\langle r_k^2\rangle = \frac{\sum_\br r_k^2 k_\br(t)}{\sum_\br k_\br(t)}
\sim 2Dt
\ .
\label{diff4}
\eeq

There is a simple scaling relation between the two-particle Green's function $K_{\br,0}$ in Eq. 
(\ref{transition00}) and saddle-point expression $c_{\br,0}$ in Eq. (\ref{corr8}) as
\beq
\sum_\br r_k^2K_{\br,0}(i\epsilon)\sim
\frac{{\bar K}}{b^2\epsilon^2}\sum_\br r_k^2c_{\br,0}
\ .
\label{scaling3}
\eeq
This result can be considered as an extension of the self-consistent Born approximation to $K_{\br,\br'}$.
\vskip0.5cm

\no
{\it Example}: Weyl fermions with random gap: $n=2$,
$H_0=i\partial_x\sigma_1+i\partial_y\sigma_2$,
$H_1=\sigma_3$, $U=\sigma_1$, $Q_1=0$, $Q_2=-i(\eta/2)\sigma_3$, where $\{\sigma_j\}$ are Pauli matrices.
The saddle-point equation reads in this case \cite{Ziegler2009}
\[
{\rm Tr}_2\left[(g_{+}g_{-})_{\br,\br}\right]=g^{-1}
\ .
\]
Inserting this in our expressions above, we obtain
$b=4i\eta/g$, ${\bar K}/b^2=-1/4$,
\[
c_{\br,0}=-4\eta^2{\rm Tr}_2\left[g_{+,0,\br}g_{-,\br,0}\right] , \ \ \
\sum_\br r_k^2K_{\br,0}(i\epsilon)\sim
-\frac{1}{4\epsilon^2}\sum_\br r_k^2c_{\br,0}
=\frac{1}{2\pi\epsilon^2}
\ .
\]
Here we have fixed the cut-off $\Lambda$ in Eq. (\ref{gf6}) such that $\det(\kappa)=1$.
The conductivity $\sigma$ can be calculated from this expression via the Kubo approach by an analytic continuation 
$\epsilon\to i\omega/2$ \cite{Ziegler2009}:
\beq
\sigma\sim -\frac{e^2}{2h}\omega^2 \sum_\br r_k^2K_{\br,0}(-\omega/2)
=\frac{e^2}{\pi h}
\ ,
\label{min_cond}
\eeq
which is the well-known minimal conductivity of graphene (except for 
an additional degeneracy factor 4) \cite{novoselov05}. The disorder 
independent conductivity reflects the wellknown fact that the conductivity
can not distinguish between ballistic and diffusive transport of Weyl fermions \cite{abergel10}. 
The diffusive behavior was also found in recent numerical simulations by Chalker et al. \cite{chalker}
and Medvedyeva et al. \cite{beenakker10}.

\subsection{Broken particle-hole symmetry}
\label{sect:5}

We introduce a chemical potential $\mu$ that shifts away from particle-hole symmetry point by $\pm\mu$
in the Hamiltonian
\beq
{\bar H}=\pmatrix{
H+\mu\sigma_0 & 0 & 0 & 0\cr
0 & H-\mu\sigma_0 & 0 & 0\cr
0 & 0 & H^T-\mu\sigma_0 & 0 \cr
0 & 0 & 0 & H^T+\mu\sigma_0 \cr
}
\ .
\label{ext_ham}
\eeq
Then we define the Green's function in analogy to ${\hat G}(i\epsilon)$ as
\beq
{\bar G}(i\epsilon)= ({\bar H}+i\epsilon)^{-1}
\ .
\label{gf0}
\eeq
The generalization of transformation matrix ${\hat S}$ in Eq. (\ref{S}) then is
\beq
{\bar S}=\pmatrix{
0 & 0 & \varphi_1U & 0 \cr
0 & 0 & 0 & \varphi_2 U \cr
\varphi_1'U^\dagger & 0 & 0 & 0 \cr
0 & \varphi_2'U^\dagger & 0 & 0\cr
}
\label{ext_transformation}
\eeq
which anticommutes with ${\bar H}$: ${\bar S}{\bar H}=-{\bar H}{\bar S}$.
This implies the symmetry transformation
\[
e^{\bar S}{\bar H}e^{\bar S}={\bar H}
\]
and ${\rm detg}(e^{\bar S})=\exp({\rm Trg}{\bar S})=1$. Now we can 
employ the expansion of Eq. (\ref{expansion2}) to obtain the nonlinear sigma model. It turns
out that the fourth-order term in ${\bar S}$ does not vanish for $\mu\ne0$, in contrast to 
the result in Eq. (\ref{nonlinear1}).

\section{Conclusions}

We have seen that the discrete particle-hole symmetry of the Hamiltonian $H \to -UH^*U^\dagger =H$ can lead to
a diffusive behavior. For this result it is crucial that no additional continuous symmetry exists for the $H$. A typical 
realization of this case are two-dimensional Weyl-Dirac fermions with random gap \cite{ziegler97}. 
The diffusive behavior requires a non-vanishing symmetry-breaking term ${\hat Q}_2$, which reflects
spontaneous breaking of the symmetry in Eq. (\ref{symm3}). ${\hat Q}_2$ must be determined as a solution
of the saddle-point equation. This can, depending on the specific Hamiltonian $H$, generate a complex phase 
diagram with metallic (i.e. diffusive), insulating and quantum-Hall phases (c.f. \cite{Ziegler2009}). 

A central fact in Sect. \ref{sect:diff} is that the saddle-point integration in Eq. (\ref{fint5}) is restricted 
to a two-component Grassmann field $(\varphi,\varphi')$. This is crucial for the derivation of the main result. 
The integration would be over a larger manifold when the underlying Hamiltonian has additional 
symmetries or in the absence of particle-hole symmetry. The latter case was briefly discussed in Sect. \ref{sect:5}
where we introduced a shift away from the particle-hole symmetry point. The integration over a larger
manifold may result in a non-diffusive behavior. 

There is a large number of publications on the subject of disordered particle-hole symmetric 
Hamiltonians (class D), which are based on (i) field theory (in particular, nonlinear sigma models), 
(ii) related network models and (iii) numerical simulations. A discussion with many references can 
be found, for instance, in Ref. \cite{evers08}. Unfortunately, there is no simple conclusion from 
all the publications because the details of the results depend on the specific form of the Hamiltonians 
or the network models, the distribution of disorder as well as on the approximations used in analytic 
treatments. Moreover, the mapping from network models onto Hamiltonian models is only understood on 
an approximative level \cite{ho96,eckern98}.

The approach discussed in this paper, which was originally proposed in Ref. \cite{ziegler97}, offers
an alternative to the nonlinear sigma model used in Ref. \cite{bocquet00}. The main difference
between the two approaches is that the former is not supersymmetric, in contrast to the latter. 
The reason is that we started from the asymmetric two-particle (Bose-Fermi) Hamiltonian 
${\hat H}=diag(H,H^T)$ in the construction of the functional integral in Eq. (\ref{action0}),
whereas Bocquet et al. used the symmetric two-particle (Bose-Fermi) Hamiltonian ${\hat H}=diag(H,H)$.
This difference has several consequences for the effective field theory of the average Green's 
functions. First, the saddle-point manifold defined in Eq. (\ref{transf3}) is different from the   
ortho-symplectic Lee group $OSp(2n|2n)/GL(n|n)$ which generates the manifold of the symmetric 
approach \cite{bocquet00}. Second, the massless mode is only the two-component Grassmann field 
$(\varphi,\varphi')$ in the asymmetric approach, whereas it consists of Grassmann and Goldstone 
(bosonic) components in the symmetric approach.
Thus the saddle-point integration is more complex in the latter. It was treated within a renormalization-group
approach, which provides an ideal metallic fixed point with infinite conductivity, in contrast to our 
finite conductivity in Eq. (\ref{min_cond}). 
Besides its technical simplicity, the asymmetric approach provides a metal-insulator phase diagram 
\cite{Ziegler2009}, which agrees qualitatively with the numerically determined phase diagram of 
Chalker et al. \cite{chalker}.

\vskip1cm
\no
Acknowledgment: I am grateful for the hospitality at the Bar-Ilan University where
part of this work was carried out during my sabbatical. Financial 
support by the DFG grant ZI 305/51 is also gratefully acknowledged.


\appendix

\section{Jacobian}
\label{app:jacobian}

The inverse Jacobian in Eq. (\ref{fint5}) reads
\beq
J^{-1}={\rm detg}\left( {\hat H}_0+i\epsilon +2{\hat Q}_1{\hat H}_1+2{\hat Q}_2{\hat H}_1e^{2{\hat S}} \right)
={\rm detg}\left({\hat H}_0+i\epsilon + 2{\hat Q}_1{\hat H}_1 -2{\hat Q}_2{\hat H}_1
+ 4{\hat Q}_2{\hat H}_1({\bf 1}-{\hat S})^{-1} \right)
\ .
\label{jacob1}
\eeq
After pulling out the factor $({\bf 1}-{\hat S})^{-1}$ we get
\[
J^{-1}
={\rm detg}({\bf 1}-{\hat S})^{-1}
{\rm detg}\left({\hat H}_0+i\epsilon + 2{\hat Q}_0{\hat H}_1 
-[{\hat H}_0+i\epsilon + 2({\hat Q}_1-{\hat Q_2}){\hat H}_1]{\hat S}\right)
\ .
\]
The (anti) commutation relation of ${\hat Q}_1$ (${\hat Q}_2$) implies
\[
iUP_j=-(-1)^jQ_jU , \ \ \ U^\dagger Q_j=-i(-1)^j P_j U^\dagger \ \ \ (j=1,2)
\]
and yields
\[
J^{-1}=\det[{\bf 1}(1-\varphi\varphi')]^{-1}\frac{\det(H_0+i\epsilon+2Q_0H_1)}{\det(H_0^T+i\epsilon+2iP_0H_1^T)}
\]
\[
\times
\det({\bf 1}-[H_0+i\epsilon+2(Q_1-Q_2)H_1]\varphi [H_0-i\epsilon+2(Q_1-Q_2)H_1]^{-1}
(H_0-i\epsilon+2Q_0H_1)\varphi'(H_0+i\epsilon+2Q_0H_1)^{-1})
\ .
\]
In the second factor, $P_j$ can be expressed by $Q_j$ such that
\[
\det(H_0^T+i\epsilon+2iP_0H_1^T)
=\det(U(H_0^T+i\epsilon+2iP_0H_1^T)U^\dagger)
=\det(-\left[H_0-i\epsilon+2(Q_1-Q_2)H_1\right])
\ .
\]
With the identities
\[
[H_0-i\epsilon+2(Q_1-Q_2)H_1]^{-1}(H_0-i\epsilon+2(Q_1+Q_2)H_1)
={\bf 1}+[H_0-i\epsilon+2(Q_1-Q_2)H_1]^{-1}4Q_2H_1
\]
\[
=:{\bf 1}+4g_-Q_2H_1
\]
and
\[
[H_0+i\epsilon+2(Q_1+Q_2)H_1]^{-1}(H_0-i\epsilon+2(Q_1+Q_2)H_1)
={\bf 1}-[H_0+i\epsilon+2(Q_1+Q_2)H_1]^{-1}4Q_2H_1
\]
\[
=:{\bf 1}-4g_+Q_2H_1
\]
and with $\det({\bf 1}-\varphi\varphi')^{-1}= \det({\bf 1}+\varphi\varphi')$ we get eventually
\[
J^{-1}=\frac{\det(H_0+i\epsilon+2Q_0H_1)}{\det(-\left[H_0-i\epsilon+2(Q_1-Q_2)H_1\right])}\det[{\bf 1}(1+\varphi\varphi')]
\]
\beq
\times
\det({\bf 1}-\{{\bf 1}-4[H_0+i\epsilon+2(Q_1+Q_2)H_1]^{-1}Q_2H_1\}\varphi 
\{{\bf 1}+4[H_0-i\epsilon+2(Q_1-Q_2)H_1]^{-1}Q_2H_1\}\varphi')
\ .
\label{jacobian5a}
\eeq
Moreover, we have
\[
\det({\bf 1}-\{{\bf 1}-4[H_0+i\epsilon+2(Q_1+Q_2)H_1]^{-1}Q_2H_1\}\varphi 
\{{\bf 1}+4[H_0-i\epsilon+2(Q_1-Q_2)H_1]^{-1}Q_2H_1\}\varphi')
\]
\[
=\det({\bf 1}-\varphi\varphi'
+4[H_0+i\epsilon+2(Q_1+Q_2)H_1]^{-1}Q_2H_1\varphi \varphi'
-4\varphi [H_0-i\epsilon+2(Q_1-Q_2)H_1]^{-1}Q_2H_1\varphi'
\]
\[
+16[H_0+i\epsilon+2(Q_1+Q_2)H_1]^{-1}Q_2H_1\varphi [H_0-i\epsilon+2(Q_1-Q_2)H_1]^{-1}Q_2H_1\varphi')
\]
\[
=\det({\bf 1}-\varphi\varphi'
+4g_+Q_2H_1\varphi \varphi'-4\varphi g_-Q_2H_1\varphi'+16g_+Q_2H_1\varphi g_-Q_2H_1\varphi')
\ .
\]
Thus, we get for the expression in Eq. (\ref{jacobian5a})
\[
J^{-1}=\frac{\det(H_0+i\epsilon+2Q_0H_1)}{\det(-\left[H_0-i\epsilon+2(Q_1-Q_2)H_1\right])}
\det({\bf 1}+4g_+Q_2H_1\varphi \varphi'-4\varphi g_-Q_2H_1\varphi'+16g_+Q_2H_1\varphi g_-Q_2H_1\varphi')
\ .
\]

\section{Scaling transformation}
\label{app:scaling}

The Green's function of the saddle-point approximation in Eq. (\ref{green4}) reads in Fourier representation
\beq
g_r=\int_0^{\Lambda}\frac{\int_0^{2\pi}e^{iqr\cos\alpha}d\alpha}{i\epsilon+m+q^2}qdq
\ ,
\label{gf6}
\eeq
where $m$ is an effective mass that is created by the saddle-point matrices $Q_1\pm Q_2$.
Rescaling $r\to\Delta r$ then gives
\beq
g_{\Delta r}
=\int_0^{\Lambda}\frac{\int_0^{2\pi}e^{i\Delta qr\cos\alpha}d\alpha}{i\epsilon+m+q^2}qdq
=\Delta^{-2}\int_0^{\Delta\Lambda}
\frac{\int_0^{2\pi}e^{ipr\cos\alpha}d\alpha}{i\epsilon+m+p^2/\Delta^2}pdp
\sim\Delta^{-2}g_r
\eeq
if $m\sim 1$, since the integral is dominated by small $p$ and does not depend on the cut-off $\Delta\Lambda$.


\end{document}